%% file: revTalk.tex
\documentclass[12pt]{article}
\usepackage{epsfig}

\input econfmacros.tex
\def\Title#1{\begin{center} {\Large {\bf #1} } \end{center}}
\begin{document}
\rightline{TECHNION-PH-12-10}
\rightline{EFI 12-17}
\rightline{UdeM-GPP-TH-12-209}
\bigskip
\Title{Direct CP Violation in $D$ Decays in view of LHCb and CDF Results
\footnote{\it The Tenth International Conference
on Flavor Physics and CP Violation - FPCP2012, May 21\\
\vskip -4mm 
~~~- 25 2012, Hefei, China.}}
%\bigskip

\begin{raggedright}  
{\it Bhubanjyoti Bhattacharya\\
Physique des Particules, Universit\'e de Montr\'eal\\
C.P. 6128, succ. centre-ville, Montr\'eal\\
Quebec, Canada H3C 3I7}\\
\bigskip
{\it Michael Gronau~\footnote{~Speaker.}\index{Gronau, M.}\\
Department of Physics\\
Technion - Israel Institute of Technology\\
Haifa 32000, Israel}\\
\bigskip
{\it Jonathan L. Rosner\\
Enrico Fermi Institute and Department of Physics\\
University of Chicago, 5620 S. Ellis Ave.\\
Chicago, Illinois 60637, USA}
\bigskip\bigskip
\end{raggedright}
\begin{quote}
The LHCb and CDF Collaborations have recently reported evidence for a CP asymmetry
around $-0.7\%$ in $\Delta A_{CP}$, the difference between $A_{CP}(D^0 \to K^+ K^-)$
and $A_{CP}(D^0 \to \pi^+\pi^-)$.  In the Standard Model this effect may be accounted for  
by enhanced $1/m_c$ corrections in a CP-violating penguin amplitude governed by a 
Cabibbo-Kobayashi-Maskawa (CKM) factor $V^*_{cb}V_{ub}$.
A consistent scheme based on broken flavor SU(3) is presented relating Cabibbo-favored 
(CF) and singly-Cabibbo-suppressed (SCS) $D$ meson decay rates into two pseudoscalars.
Two important ingredients supporting the above interpretation for $\Delta A_{CP}$  
are a large exchange amplitude in 
CF decays which is formally $1/m_c$-suppressed, and a pure $\Delta U=0$ 
U-spin breaking CP-conserving penguin amplitude involving $V^*_{cs}V_{us}$ and 
$V^*_{cd}V_{ud}$ which accounts for the difference
between the $D^0\to K^+K^-$ and $D^0\to \pi^+\pi^-$ decay rates. 
The magnitudes of the CP conserving and CP violating penguin 
amplitudes, where the former involves U-spin breaking at a level of $10\%$,  
are shown to be related to each other by the magnitudes of corresponding
CKM factors. This simple scheme leads to preferable sign predictions for CP 
asymmetries in charmed meson decays into two pseudoscalars and to correlations 
between asymmetries in two pairs of these processes.
\end{quote}

\section{Introduction}

CP violation was first observed at a level of $10^{-3}$ in kaon decays. A variety of CP 
asymmetries have been measured subsequently at the tens of percent level in $B$ 
decays~\cite{Trabelsi,Schiller}. The Standard Model, which accounts adequately for  all these unique observations~\cite{Gronau:2007xg}, has often been used to claim predicting naturally 
very small CP asymmetries in SCS decays of charmed 
particles, of order $10^{-3}$ or less
\cite{Buccella:1994nf,Bianco:2003vb,Grossman:2006jg,Bigi:2011re}.  
These decays, dominated by physics of the first two quark families, involve a tiny 
CP-violating penguin contribution of the third family ($P_b$) suppressed both  by 
the smallness of the 
CKM matrix element  $V^*_{cb}V_{ub}$ and by the relatively 
small $b$ quark mass (compared to a large penguin amplitude in $B$ decays involving 
$V^*_{tb}V_{ts}$ and the heavy $t$ quark mass). 
However, a very early study has proposed that the penguin 
amplitude in SCS $D$ decays may be enhanced by nonperturbative effects in analogy to the
$s \to d$ penguin amplitude in $K \to \pi \pi$~~\cite{Golden:1989qx}. 

During the first half of 2012 the LHCb and CDF Collaborations have reported 
independently evidence for CP-violating charm decays in the difference $\Delta A_{CP}$ 
 between CP asymmetries in $D^0 \to K^+ K^-$ and $D^0 \to \pi^+ \pi^-$. The averaged value
 of the two asymmetry measurements was found  to be nonzero at $\sim 4\sigma$ at the somewhat 
 unexpected high level of about 
 $\Delta A_{CP} \simeq -0.7\%$~\cite{Aaij:2011in,CDF,Vagnoni}. These results have led to a continuous  flood of theory papers, 
 some trying to recover these asymmetries within the Standard 
 Model~\cite{Bigi:2011em}--\cite{Cheng:2012xb},
 %Isidori:2011qw,Brod:2011re}, \cite{Pirtskhalava:2011va}-
 while others offering also interpretations beyond the Standard Model
 \cite{Bigi:2011em,Isidori:2011qw}, \cite{Wang:2011uu}--\cite{Chen:2012}.
%Rozanov:2011gj,Hochberg:2011ru,Giudice:2012qq,
%Altmannshofer:2012ur,Chen:2012am,Feldmann:2012js,Hiller:2012wf,Grossman:2012eb,
%Mannel:2012hb
 
 Motivated by the LHCb and CDF experimental results described in Sec.\ 2, we will 
 start this review by  discussing in Sec.\ 3 a crude estimate of $\Delta A_{CP}$ within the 
 Standard Model.  We point out an intrinsic uncertainty in the estimated asymmetry 
 from $1/m_c$ corrections, 
 presenting in Sec.\ 4 experimental evidence for such large corrections in decay rates for 
 CF $D$ decays. In Sec.\ 5 we describe a scheme~\cite{Bhattacharya:2012ah} based 
 on broken flavor SU(3)
 relating decay rates for CF and SCS charmed meson decays, which accounts consistently 
 and quite precisely for both CF and SCS decay rates of $D$ mesons into two 
 pseudoscalars, explaining in particular the different decay rates for $D^0\to K^+K^-$ 
 and $D^0\to \pi^+\pi^-$.  We introduce two types of U-spin breaking in SCS decays: 
 (1) U-spin breaking in $\Delta U=1$ tree amplitudes given in terms of ratios of light 
 pseudoscalar meson decay constants and ratios of $D$ meson form factors.  
 (2) A pure U-spin breaking $\Delta U=0$ penguin amplitude.    
In a short Sec.\ 6 we show that a $10\%$ U-spin breaking in this penguin amplitude, 
required to account for SCS decay rates, can explain naturally the measured CP asymmetry
$\Delta A_{CP}$ in terms of a penguin amplitude $P_b$ involving $V^*_{cb}V_{ub}$.  In this
framework all CP asymmetries in SCS $D$ decays into two pseudocalars depend
on two parameters, the magnitude and strong phase of  $P_b$.
Consequently one expects correlations between asymmetries in different processes. 
We discuss such correlations in Sec.\ 7 and conclude in Sec.\ 8.

\section{LHCb and CDF asymmetry measurements}

One defines a CP asymmetry in $D\to f$:
\beq\label{A_CP}
A_{CP}(f) \equiv \frac{\Gamma(D \to f) - \Gamma(\bar D \to \bar f)}
                      {\Gamma(D \to f) + \Gamma(\bar D \to \bar f)}~.
\eeqn
For a CP-eigenstate, such as $f=\pi^+\pi^-$ or $K^+K^-$,
\beqa
A_{CP}(f_{CP}) & \equiv & \frac{\Gamma(D^0 \to f_{CP}) -
 \Gamma(\bar D^0 \to f_{CP})}{\Gamma(D^0 \to f_{CP}) + \Gamma(\bar D^0 \to f_{CP})}
 \nonumber\\
& \approx & A_{CP}^{\rm dir}(f) + \frac{\langle t(f)\rangle}{\tau_D}A_{CP}^{\rm ind}(f)~,
\eeqan
where $t$ is the proper decay time and $\tau_D$ is the $D$ meson lifetime.

The indirect asymmetry $ A_{CP}^{\rm dir}$ induced by $D^0$-$\bar D^0$ mixing 
($q/p$) is approximately
\beq
A_{CP}^{\rm ind}(f) \sim x\sin\phi~~(x\equiv \frac{\Delta m}{\Gamma}),~~~
\phi\equiv {\rm arg}\left[(q/p)(\bar A_f/A_f)\right]~.
\eeqn
Thus $A_{CP}^{\rm ind}(f)$ is approximately independent of $f$ because $A_f$
and $\bar A_f$ involves a tiny weak phase difference.

One now defines the difference of two asymmetries:
\beq
\Delta A_{CP}  \equiv  A_{CP}(K^+ K^-) - A_{CP}(\pi^+ \pi^-)
\approx  \Delta A_{CP}^{\rm dir} + \frac{\Delta\langle t\rangle}{\tau_D}A_{CP}^{\rm ind}~,
\eeqn
where 
\beq
\Delta\langle t\rangle\equiv \langle t(K^+K^-)\rangle - \langle t(\pi^+\pi^-)\rangle~.
\eeqn

The measurements of LHCb and CDF are:\\\\
LHCb~\cite{Aaij:2011in}
\beqa 
\Delta A_{CP}  & = & [-0.82 \pm 0.21({\rm stat}) \pm 0.11({\rm syst})]\%~,
\nonumber\\
\frac{\Delta\langle t \rangle}{\tau_D} & = & [9.83\pm 0.22 ({\rm stat}) \pm 0.19 ({\rm syst})]\%~,
\eeqan
%MG updated CDF Ref
CDF~\cite{CDF}
\beqa
\Delta A_{CP} & = & [-0.62 \pm 0.21 ({\rm stat}) \pm 0.10 ({\rm syst})]\%~,
\nonumber\\
%MG replaced 26 by 27
\frac{\Delta\langle t \rangle}{\tau_D} &  = & [27 \pm 1]\%~.
\eeqan
Combining the two results and assuming uncorrelated Gaussian errors, one finds 
averages~\cite{average}:
\beq\label{AveAsym}
\Delta A_{CP}^{\rm dir} = (-0.656 \pm 0.154)\%~~,~~~~
A_{CP}^{\rm ind} = (-0.025 \pm 0.231)\%~.
\eeqn
Thus the direct asymmetry differs from zero by about $4\sigma$, while the indirect asymmetry is consistent with zero. We note that earlier asymmetry measurements by the E687~\cite{E687}, E791~\cite{E791}, FOCUS~\cite{FOCUS}, CLEO~\cite{CLEO}, BaBar~\cite{BaBar} and 
Belle~\cite{Belle} collaborations have a negligible effect on these averages.

We will also use results obtained by CDF for separate 
asymmetries in $D^0\to K^+K^-$ and $D^0\to \pi^+\pi^-$~\cite{Aaltonen:2011se},
\beqa\label{BoundsAsym}
A_{CP}(D^0 \to K^+ K^-) & = & (-0.24 \pm 0.22 \pm 0.09)\%,~\nonumber\\
A_{CP}(D^0 \to \pi^+ \pi^-) & = & (0.22 \pm 0.24 \pm 0.11)\%~,
\eeqan
for which we calculate the corresponding 90\% confidence level limits:
\beq \label{eqn:limits}
-0.63\% \le A_{CP}(D^0 \to K^+ K^-) \le 0.15\%~,~~
-0.21\% \le A_{CP}(D^0 \to \pi^+ \pi^-) \le 0.65\%~.
\eeqn

\section{Estimate of $\Delta A_{CP}$ in the Standard Model}

SCS charmed meson decays are described by an effective weak Hamiltonian
$H_{\rm eff}$ at a scale $\mu \sim m_c$:
\beq\label{Heff}
H_{\rm eff}^{\rm SCS} = \frac{G_F}{\sqrt2}\left [\Sigma_{i=d,s}V^*_{ci}V_{ui}
(C_1Q^i_1 + C_2Q^i_2) - V^*_{cb}V_{ub}\Sigma_{j=3}^6 C_jQ_j + C_{8g}Q_{8g}\right] 
+ h.c.
 \eeqn
 The $(V-A)\times (V-A)$ current-current ``tree" operators  $Q_{1,2}^i$ have flavor 
 structure $(\bar i c)(\bar u i)$
 where $i=d, s$, while the penguin operators $Q_{3,..,6}$ have a structure 
 $(\bar u c)\Sigma_q(\bar q q)$,
 where $q=u,d,s$. The Wilson coefficients can be evaluated in perturbation theory at 
 $\mu \sim m_c$. In contrast, it is extremely difficult to calculate the hadronic matrix elements of these operators and of the chromomagnetic operator $Q_{8g}\equiv -\frac{g_s}{8\pi^2}m_c\bar u\sigma_{\mu\nu}(1+\gamma_5)G^{\mu\nu}c$.
 
 An estimate of the dominant contribution to $\Delta A_{CP}$, which originates in the interference 
 of tree and penguin amplitudes,  requires calculating the ratio $r\equiv |P_b|/|T|$ of 
 penguin and tree amplitudes for $D^0\to \pi^+\pi^-$ and $D^0\to K^+K^-$. 
 Denoting by $\delta_{\pi\pi}$ the strong phase difference between these amplitudes in 
 $D^0\to \pi^+\pi^-$, one has
 \beq
 A_{CP}(D^0\to\pi^+\pi^-) \approx 2r\sin\delta_{\pi\pi}\sin\gamma~.
 \eeqn
 As we will show in Sec. 5, the asymmetries in $D^0\to \pi^+\pi^-$ and $D^0\to K^+K^-$ are 
 related by an approximate U-spin symmetry,
 \beq\label{ACP-Uspin}
 A_{CP}(D^0\to K^+K^-) \simeq -A_{CP}(D^0\to\pi^+\pi^-)~,
 \eeqn
  thus
  \beq
  \Delta A_{CP} \simeq - 4r\sin\delta_{\pi\pi}\sin\gamma~.
  \eeqn
  The weak phase $\gamma$ is known to be around $70^\circ$~\cite{Charles:2011va} 
  with $\sin\gamma\simeq 1$. One may assume for simplicity $\sin\delta _{\pi\pi}\sim 1$, 
  hence $\Delta A_{CP} \sim -4r$. 
   The LHCb and CDF averaged direct asymmetry in Eq.~(\ref{AveAsym}) correspond to
  $r\sim(17\pm 4)\times 10^{-4}$. 
 Factoring out CKM matrix elements in $P_b$ and $T$ by defining 
 $P_b \equiv V^*_{cb}V_{ub}\,{\cal P}_b$ and $T = V^*_{cd(s)}V_{ud(s)}\,{\cal T}$, 
 one has~\cite{PDG}:
 \beq\label{r-exp}
 r \equiv \frac {|P_b|}{|T |}
 = \frac{|V^*_{cb}V_{ub}|}{|V^*_{cd}V_{ud}|}\frac{|{{\cal P}_b}|}
 {|{\cal T}|} \simeq 7\times 10^{-4}\frac{|{{\cal P}_b}|}{| {\cal T}|}~~~~ {\rm implying}~~~~ 
 \frac{|{{\cal P}_b}|}{| {\cal T}|} 
 \sim 2 - 3~.
 \eeqn

 The question is: are we able to provide a reliable estimate for $|{{\cal P}_b}|/
 |{\cal T}|$, the ratio of reduced penguin and tree matrix elements?
 
 A naive estimate based on perturbation theory would suggest $|{{\cal P}_b}|/
 | {\cal T}|\sim \alpha_s(m_c)/\pi$, lying an order of magnitude below (\ref{r-exp}).
However, as $m_c$ is not much larger than $\Lambda_{\rm QCD}$, one expects large 
$1/m_c$ corrections to this leading order estimate. The most direct evidence for such 
large $1/m_c$ corrections 
is provided by a large exchange amplitude contributing to CF $D$ decay rates to 
be discussed in the next section. 

Detailed calculations of $1/m_c$ corrections in $r$ are quite challenging and involve subtle sizable uncertainties.
Ref.~\cite{Brod:2011re} shows, for instance, that insertion of penguin operators into power-suppressed annihilation amplitudes, accompanied by penguin contraction matrix elements of tree operators (which are also formally $1/m_c$ suppressed) are enhanced relative to the above leading order estimate. Thus Ref.~\cite{Brod:2011re} concludes that the measured 
asymmetry $\Delta A_{CP}$ can be accommodated within the Standard Model.

Other recent attempts for calculating $r$ using somewhat different approaches
demonstrate the difficulty of this calculation. Ref.~\cite{Li:2012cfa} assumes factorization 
for parametrizing nonperturbative hadronic matrix elements of penguin operators,
thus obtaining an asymmetry considerably smaller than measured.
Ref.~\cite{Franco:2012ck} applies an oversimplified two (or three) channel  coupled S 
matrix for $\pi\pi$ and $KK$ final states, assuming SU(3) breaking of order one 
in $D^0\to K^0\bar K^0$ (in contrast, we show that SU(3) breaking of order $10\%$ in an
enhanced CP-conserving penguin amplitude suffices), thereby obtaining an asymmetry 
which is marginally consistent with the $\Delta A_{CP}$ measurement.

Thus we conclude conservatively that the seemingly large value measured for 
$\Delta A_{CP}$ can be accounted 
for by penguin enhancement and is not inconsistent with the Standard Model. This situation, anticipated many years ago in Ref.~\cite{Golden:1989qx}, resembles the situation of the 
observed $\Delta I=1/2$ enhancement in $K\to \pi\pi$, which originates to a large extent from
penguin dominance~\cite{Vainshtein:1975sv}. Both effects involve nonperturbative
uncertainties. 

In the next two sections we show evidence both for large $1/m_c$-suppressed exchange
amplitudes contributing to CF decay rates, and for enhanced CP-conserving penguin 
amplitudes contributing to SCS $D$ decay rates. Subsequently we  present in this scheme 
CP asymmetry predictions for a whole class of SCS nonstrange charmed meson decays 
into two light pseudoscalars. 

\section{CF $D$ decays into two pseudoscalars}

 CF $D$ meson decays into two pseudoscalars are conveniently described
 in terms of four types of SU(3) flavor-topology  matrix elements of 
 tree operators, $T$ (color-favored), $C$ (color-suppressed), 
 $E$ (exchange), and $A$ (annihilation)
 \cite{Gronau:1994rj,Bhattacharya:2008ss,Bhattacharya:2009ps,Cheng:2010ry}. 
 This description, which relates  amplitudes of CF decays to those of SCS decays (see
 next section),  is equivalent 
 to group theoretical expressions in terms of SU(3) reduced amplitudes. (See
 \cite{Golden:1989qx,Pirtskhalava:2011va} and references therein.)
 A fit to decay rates performed in Ref.~\cite{Bhattacharya:2009ps} using
the kinematic relation  between decay amplitudes and decay rates, 
$\Gamma = |A|^2p_{\rm cm}/8\pi M^2_D$, leads to results given in Table \ref{tab:CF}.
 %This is Table 1
 \begin{table}
\begin{center}
\begin{tabular}{c c c c} \hline 
Decay & Amplitude     & Theoretical $\b~\%$ & Experimental $\b~\%$~\cite{PDG} \\ \hline
$D^0\to K^-\pi^+$ & $T + E$  & $3.90$ & $3.88 \pm 0.05$ \\
$D^0\to \bar K^0\pi^0$ & $(C-E)/\sqrt2$ & $2.35$ & $2.38 \pm 0.08$ \\
$D^0 \to \bar K^0\eta$ &  $C/\sqrt3$ & $1.00$ & $0.96\pm 0.06$ \\
$D^0 \to \bar K^0\eta'$ &  $-(C+3E)/\sqrt6$ & $1.92$ & $1.88\pm 0.10$ \\
$D^+\to \bar K^0\pi^+$ & $T + C$  & $3.09$ & $2.93 \pm 0.09$\\
$D^+_s\to\bar K^0K^+$ & $C + A$  & $2.94$ & $2.96 \pm 0.16$\\
$D^+_s \to \pi^+\eta$ & $(T - 2A)/\sqrt3$  & $1.81$ & $1.83 \pm 0.15$\\
$D^+_s \to \pi^+\eta'$ & $2(T + A)/\sqrt6$  & $3.60$ & $3.94 \pm 0.33$\\
\hline  
\end{tabular}
\caption{Amplitude representations and comparison of experimental and fit 
branching ratios for CF decays of charmed mesons to two pseudoscalars.}
\label{tab:CF} 
\end{center}
\end{table}
The rather good quality fit, involving a slight dependence on $\eta$-$\eta'$ mixing (the values in
Table \ref{tab:CF} correspond to $\theta_\eta=19.5^\circ$), obtains the following complex 
amplitudes in units of $10^{-6}$ GeV:
\beq\label{TCEA}
T = 2.93\,,~~~C = 2.34\,e^{-i152^\circ}\,,~~~E = 1.57\,e^{i121^\circ}\,,~~~ A = 0.33\,e^{i170^\circ}~.
\eeqn 

Five comments are in order when considering CF amplitudes:
\begin{itemize}
\item The amplitude $E$ is formally suppressed by $1/m_c$ relative to $T$. 
The experimental ratio $|E|/|T| > 0.5$ implies large $1/m_c$ corrections in $E$.
This point has been alluded to in the preceding section, providing a basis
for expecting large $1/m_c$ corrections also in SCS decays.
\item The four topolgical amplitudes involve large relative strong phases.
These phases originate in final state interactions which are at least partially due to 
nearby resonances~\cite{Gronau:1999zt}.
\item CF decays involve a single CKM factor $V^*_{cs}V_{ud}$ and obtain no 
contribution from a penguin amplitude. Thus one expects $A_{CP}=0$ in these 
decays.
\item The effective weak Hamiltonian inducing CF decays has flavor structure
$( \bar s  c)(\bar u d)$ and is $\Delta U=1$.
U-spin is an SU(2) subgroup of flavor SU(3), under which $(d, s)$ transforms like 
a doublet while $u$ and $c$ are singlets.
\item CF and doubly-Cabibbo-suppressed $D$ decays (not discussed in this review)
are related to each 
other by a U-spin transformation, $d\leftrightarrow s$~\cite{Gronau:2000ru}.
\end{itemize}

\section{SCS $D$  decays into two pseudoscalars}

The effective weak Hamiltonian for SCS decays involves current-current (tree)
operators with  flavor structure  
\beq\label{HeffSCS}
H^{\rm SCS}_{\rm tree} \propto \lambda[(\bar s c_)
(\bar u s) - (\bar d c)(\bar u d)]~,
\eeqn
where $\lambda \equiv \tan\theta_{\rm Cabibbo}= 0.2317$.
We have neglected a small weak phase difference between 
$V^*_{cs}V_{us}$ and $-V^*_{cd}V_{ud}$. In fact, unitarity of the CKM matrix 
 \beq\label{small-phase}
 V^*_{cd}V_{ud} + V^*_{cs}V_{us} + V^*_{cb}V_{ub}=0~~~{\rm implies}~~
 {\rm Arg}\left (\frac{V^*_{cs}V_{us}}{-V^*_{cd}V_{ud}}\right) \approx  
\frac{|V_{cb}||V_{ub}|}{|V_{cd}||V_{ud}|}\sin\gamma \simeq 7\times 10^{-4}~.
\eeqn
We will argue in Sec.\,6 that the effect 
of this tiny phase on CP asymmetries in SCS decays is negligible.

Within flavor SU(3) the amplitudes contributing to SCS processes are
the four  tree amplitudes in CF decays, $T, C, E$ and $A$ multiplied by 
$V^*_{cs}V_{us}/V^*_{cs}V_{ud} = \lambda$ or $V^*_{cd}V_{ud}/V^*_{cs}V_{ud} 
= - \lambda$. The operator $H^{\rm SCS}_{\rm tree}$ transforms like $\Delta U=1$;
 hence in the U-spin symmetry limit (using the transformation $d\leftrightarrow s$) one has 
\beq\label{T+E}
A_{\rm tree}(D^0\to K^+K^-) = -A_{\rm tree}(D^0\to\pi^+\pi^-) = \lambda(T + E)~.
\eeqn
 
 The tiny CP-violating penguin operator in Eq.~(\ref{Heff}), of flavor structure 
 $(\bar u c)\Sigma_q(\bar q q)$, is a U-spin singlet. Therefore in the U-spin symmetry limit 
 \beq
 A_{\rm penguin}(D^0\to K^+K^-) = A_{\rm penguin}(D^0\to\pi^+\pi^-)~,
 \eeqn
 implying 
 \beq
 A_{CP}(D^0\to K^+K^-) = - A_{CP}(D^0\to \pi^+\pi^-)~, 
 \eeqn
 as has already been mentioned in Eq.~(\ref{ACP-Uspin}).
 
 We now discuss U-spin symmetry breaking which plays an important role in 
 SCS decays. The two tree amplitudes measured for $D^0\to\pi^+\pi^-$ and $D^0\to K^+K^-$,
 which are equal in the U spin symmetry limit, seem to be experimentally quite different in 
 magnitude (we neglect the tiny penguin contribution $P_b$)~\cite{PDG},
 \beq\label{ratio}
 \frac{|A(D^0\to K^+K^-)|}{|A(D^0\to\pi^+\pi^-)|} = 1.81 \pm 0.04~.
 \eeqn
 Typical U-spin breaking [or SU(3) breaking], such as in $f_K/f_\pi$ or in 
 $F(D\to K)/F(D\to \pi$), is at most of order $0.2 - 0.3$.
 The ratio (\ref{ratio}) may combine several sources of U-spin breaking.
 U-spin breaking in factorized (color-favored)  tree amplitudes~\cite{Bauer:1986bm,Neubert:1997uc}
 is given in terms of a product of ratios 
 of light meson decay constants and $D$ meson form factors
 \cite{decay-constants,Besson:2009uv} (neglecting the contribution of 
 $f_-(q^2)$ at $q^2 = m^2_{\pi, K}$):
 \beq\label{SU3tree}
 \frac{f_K}{f_\pi}\frac{f_+(D\to K)(m^2_K)}{f_+(D\to\pi)(m^2_\pi)} = 1.38~.
 \eeqn
 This factor is expected to affect the ratio $|A_{\rm tree}(D^0\to K^+K^-)|/
 |A_{\rm tree}(D^0\to\pi^+\pi^-)|$ through the contributions of $T$ but not of $E$.
 [See Eq.~(\ref{T+E}).] Thus the ratio of $\Delta U=1$ tree contributions is smaller 
 than 1.38 and is insufficient by itself to account for the measured ratio of
 amplitudes (\ref{ratio}).
 
 This situation requires another contribution to U-spin breaking from $\Delta U=0$ 
 operators. Indeed, using flavor topology amplitudes such a contribution is naturally
 provided by $c \to u$ penguin amplitudes involving $d$ and $s$ quarks with 
 different masses in the intermediate state~\cite{Bhattacharya:2012ah}. We note that these 
  amplitudes involve the same CKM factor $\lambda$ as the $\Delta U=1$ tree
  amplitudes, thereby affecting decay rates without leading to CP asymmetries. 
  (As mentioned, we neglect a tiny phase difference between $V^*_{cs}V_{us}$ and
  $-V^*_{cd}V_{ud}$; its negligible effect on CP asymmetries will be discussed in the next 
  section.) 
  These penguin amplitudes are expected naturally to increase $|A(D^0\to K^+K^-)|$ and 
  decrease $|A(D^0\to \pi^+\pi^-)|$ as required, because they contribute equally with 
  equal signs to these amplitudes, while tree amplitudes contributions involve opposite 
  signs. [See Eq. (\ref{T+E}).]
  
 We denote these pure U-spin breaking $\Delta U=0$ contributions by $P\equiv P_s-P_d$ 
 and $PA\equiv PA_s - PA_d$, representing differences between $s$ and $d$ contributions from
 penguin and penguin annihilation topologies. In order to include SU(3) 
  breaking in tree  amplitudes we define
 \beqa
T_\pi & = & T\,\cdot\,\frac{|f_{+(D^0\to\pi^-)}(m^2_\pi)|}{|f_{+(D^0\to K^-)}
(m^2_\pi)|} \,\cdot\,\frac{m^2_D - m^2_\pi}{m^2_D - m^2_K}, \\
T_K   & = & T\,\cdot\,\frac{|f_{+(D^0\to  K^-)}(m^2_K)|}{|f_{+(D^0\to K^-)}
(m^2_\pi)|} \,\cdot\,\frac{f_K}{f_\pi}.
\eeqan 
 %
  %This is Table 2
  \begin{table}
\begin{center}
\begin{tabular}{c c c c c} \hline \hline
Decay & Amplitude      & \multicolumn{2}{c}{$|A_f|$ ($10^{-7}$ GeV)} & $\phi^f_T$\\ \cline{3-4}
 Mode & representation $A_f$ & Experiment & Theory & degrees \\ \hline
$D^0\to\pi^+\pi^-$ & $-\lambda\,(T_\pi + E) + (P + PA)$  
& 4.70 $\pm$ 0.08 & 4.70 & --158.5 \\
$D^0\to K^+  K^-$ & $\lambda\,(T_K + E) + (P + PA)$ & 8.49 $\pm$ 0.10 & 8.48 & 32.5 \\
$D^0\to \pi^0\pi^0$ & $-\lambda\,(C - E)/\sqrt2 - (P + PA)/\sqrt2 $& 3.51 $\pm$ 0.11& 3.51
& 60.0 \\ 
$D^0\to K^0\bar K^0$    &$-(P +  PA) + P$  &2.39 $\pm$ 0.14 & 2.37 & --145.6 \\
$D^+\to \pi^+\pi^0$ & $-\lambda\,(T_\pi + C)/\sqrt2$ & 2.66 $\pm$ 0.07 & 2.26 & 126.3 \\ 
$D^+\to K^+\bar K^0$  & $\lambda\,(T_K - A) + P$ & 6.55 $\pm$ 0.12 & 6.87 &--4.2 \\
\hline \hline
\end{tabular}
\caption{Amplitude representations quoting strong phases; comparison of experimental
and fit branching ratios for SCS decays of charmed mesons to two pseudoscalars.}
\label{tab:SCS} 
\end{center}
\end{table}
 
 Amplitude representations $A_f$ for six SCS decays of nonstrange charmed mesons into two 
 pseudoscalars, and strong phases $\phi^f_T$ of these amplitudes calculated with respect to $T$,
 are listed in Table~\ref{tab:SCS}. Input values for tree amplitudes, $T, C, E$ and $A$ involving 
 strong phases were obtained in Eq.~(\ref{TCEA}) using CF decay rates. The
 experimental decay rates (or the quoted measured magnitudes of amplitudes $A_f$) for 
 the first three SCS processes in Table \ref{tab:SCS} are used to fit the complex
 amplitude $P + PA$. 
 
 The extracted value~\cite{Bhattacharya:2012ah}, 
 \beq\label{P+PA}
 P + PA = (0.44 + 1.41\,i)\,(10^{-7}~{\rm GeV})\,,~~{\rm with~magnitude}~
 |P + PA| = 1.48\times (10^{-7}~{\rm GeV})~,
 \eeqn
 is seen to fit the 
 data excellently, corresponding to a very low $\chi^2$ for a single degree of freedom. 
 In particular, using CF decay rates as input, this U-spin-breaking amplitude accounts 
 very well for the quite different decay rates of  $D^0\to\pi^+\pi^-$ and $D^0\to K^+K^-$, 
 which seemed {\it ab initio} like a puzzle. It also fits well the measured decay rate for 
 $D^0\to\pi^0\pi^0$.
 
 We note in passing that the ratio of tree contributions to the two 
 amplitudes for $D^0\to\pi^+\pi^-$ and $D^0\to K^+K^-$~\cite{Bhattacharya:2012ah},  
  \beqa\label{T}
 |A_{\rm tree}(D^0\to\pi^+\pi^-)| & = &  \lambda|T_\pi + E| = 5.73\times (10^{-7}~{\rm GeV})~,
 \nonumber\\
 ~~~|A_{\rm tree}(D^0\to K^+K^-)| & = & \lambda|T_K + E| = 7.42\times (10^{-7}~{\rm GeV})~,
 \eeqan
 involves U-spin breaking of $29\%$ which, as anticipated,  is indeed smaller than U-spin
 breaking in the ratio (\ref{SU3tree}).
 
 The approximately real extracted value of $P$, $P = (-1.52 + 0.08\,i)\,(10^{-7}$ GeV),
 comparable in magnitude to $|P + PA|$ but involving a 
 large error, is obtained using rate measurements for $D^0\to K^0 \bar K^0$ and 
 $D^+\to K^+\bar K^0$~\cite{Bhattacharya:2012ah}.  We note that while all four $D^0$ decay 
 rates are fitted excellently, the fit to the two $D^+$ decay rates is of only moderate quality.
 
 \section{Estimating U-spin breaking in $P+PA$ from $P_b$}
 
 Both the CP-conserving penguin amplitude, $P+PA$ or $P$ (which 
 are of comparable magnitudes and are purely U-spin breaking), and the CP-violating penguin amplitudes $P_b$ are of nonperturbative nature. Assuming that their enhancement has 
 a common origin due to $1/m_c$ corrections, we will estimate the level of U-spin breaking 
 in $P + PA$. (See also discussion  in Ref.~\cite{Brod:2012ud}.)
 
 As mentioned, these two amplitudes involve different
 CKM factors, $V^*_{cs}V_{us}$ (or $V^*_{cd}V_{ud}$) and $V^*_{cb}V_{ub}$, respectively,
 which may be factored out  defining $P + PA \equiv V^*_{cs}V_{us}\,({\cal P + PA})$
 and $P_b \equiv V^*_{cb}V_{ub}\,{\cal P}_b$. [See also line above Eq.~(\ref{r-exp})].
 The ratio 
 \beq
 |{\cal P + PA}|/|{\cal T}| =  |P + PA|/|T|= 0.20 - 0.26~,
 \eeqn 
 is calculated from Eq.~(\ref{P+PA}) and (\ref{T}), where $T$ is the tree amplitude 
 in $D^0\to \pi^+\pi^-$ or $D^0\to K^+K^-$. (These two tree amplitudes differ by about 
 $30\%$.) A ratio $|{\cal P}_b|/|{\cal T}|= 2 - 3$ was calculated in Eq.~(\ref{r-exp}) using  the 
  average value of $\Delta A_{CP}$ measured by LHCb and CDF and assuming a large strong
  phase difference $\delta$ between ${\cal P}_b$ and ${\cal T}$. Taking the ratio 
  of these two ratios we conclude:
  \beq\label{SU3}
  \frac{|{\cal P + PA}|}{|{\cal P}_b|} \simeq 0.10~.
  \eeqn
  This value provides an estimate for U-spin breaking in $P + PA$,
  \beq\label{SU3P+PA}
  \frac{|(P_s + PA_s) - (P_d + PA_d)|}{|P_s + PA_s|} \sim 0. 10~,
  \eeqn
  which seems quite reasonable.
  
  We now discuss briefly the effect on CP asymmetries in SCS decays of a small weak 
  phase difference between $V^*_{cs}V_{us}$ and $-V^*_{cd}V_{ud}$ 
  ($\simeq 7\times 10^{-4}$) which has been neglected in Eq.\,(\ref{HeffSCS}). 
  Using unitarity of the CKM matrix one may rewrite $P$ and $PA$ in an explicitly U-spin 
  breaking form,
  \beq
  P = V^*_{cs}V_{us}({\cal P}_s-{\cal P}_d)~,~~
  PA = V^*_{cs}V_{us}({\cal PA}_s-{\cal PA}_d)~,
  \eeqn
  with $P_b=V^*_{cb}V_{ub}({\cal P}_b - {\cal P}_d)$, 
  $PA_b=V^*_{cb}V_{ub}({\cal PA}_b - {\cal PA}_d)$ . In the next section  we will assume
 that $PA_b$ is negligible relative to $P_b$, proposing tests for this assumption. 
 
 Interference of tree and $P + PA$ amplitudes in $D^0\to K^+K^-$ does not contribute to the asymmetry in this process because the two amplitudes involve a common CKM factor 
 $V^*_{cs}V_{us}$ and a common weak phase.
 On the other hand, interference of corresponding amplitudes in $D^0\to\pi^+\pi^-$
 involves the above small phase difference, thus leading to a nonzero contribution
 to  $A_{CP}(D^0\to\pi^+\pi^-)$. Using information about 
 magnitudes and strong phases of these amplitudes, we calculate this contribution to 
 be at a negligible level of $1\times10^{-4}$. A contribution at the same level is calculated for 
 $A_{CP}(D^+\to K^+\bar K^0)$ from interference of  $T_K$ and $A$ amplitudes also
 involving this tiny weak phase difference. 
      
  \section{Predicting CP asymmetries in SCS decays} 
 
Assuming flavor SU(3) for the CP-violating amplitude $P_b$ [SU(3) breaking in the 
corresponding CP-conserving amplitude $P+PA$ has been shown to be only about 
$10\%$], we now include
$P_b$ in the six amplitudes in Table \ref{tab:SCS}. The resulting complete amplitude 
representations are given in Table \ref{tab:SCSincPb}.

Denoting $P_b \equiv |P_b|\,e^ {i(\delta - \gamma)}$ the amplitude for $D\to f$ may be written as
\beq
{\cal A}(D \to f) = |A_f|e^{i\phi^f_T}\left (1 + \frac{|P_b|}{|A_f|}\,e^{i(\delta-\phi^f_T -
\gamma)}\right)~,
\eeqn
where $|A_f|$ and $\phi^f_T$ were defined and calculated in Table \ref{tab:SCS}.
The amplitude for $\bar D \to \bar f$ is obtained from this expression by $-\gamma 
\to \gamma$. Using the definition (\ref{A_CP}) one obtains CP asymmetries
\beqa\label{CPAsym}
A_{CP}(f) & = & \frac{2(|P_b|/|A_f|)\sin\gamma\sin(\delta - \phi^f_T)}
{1 + (|P_b|/|A_f|)^2 + 2(P_b|/|A_f|)\cos\gamma\cos(\delta - \phi^f_T)}
\nonumber\\
& \approx & 2(|P_b|/|A_f|)\sin\gamma\sin(\delta - \phi^f_T)~,
\eeqan
where terms quadratic in $|P_b|/|A_f|$ have been neglected in the second line. 

%This is  Talbe 3
  \begin{table}
\begin{center}
\begin{tabular}{c c } \hline \hline
Decay Mode & Amplitude  representation    \\ \hline
$D^0\to\pi^+\pi^-$ & $-\lambda\,(T_\pi + E) + (P + PA) + P_b$  \\
$D^0\to K^+  K^-$ & $\lambda\,(T_K + E) + (P + PA) + P_b$  \\
$D^0\to \pi^0\pi^0$ & $-\lambda\,(C - E)/\sqrt2 - (P + PA)/\sqrt2 -P_b/\sqrt2$ \\ 
$D^0\to K^0\bar K^0$    &$-(P +  PA) + P$  \\
$D^+\to \pi^+\pi^0$ & $-\lambda\,(T_\pi + C)/\sqrt2$  \\ 
$D^+\to K^+\bar K^0$  & $\lambda\,(T_K - A) + P +P_b$  \\
\hline \hline
\end{tabular}
\caption{Amplitude representations including $P_b$ for SCS decays of charmed mesons to two pseudoscalars.}
\label{tab:SCSincPb} 
\end{center}
\end{table}
%
%This is Fig. 1
\begin{figure}
\begin{center}
\includegraphics[width=0.7\textwidth]{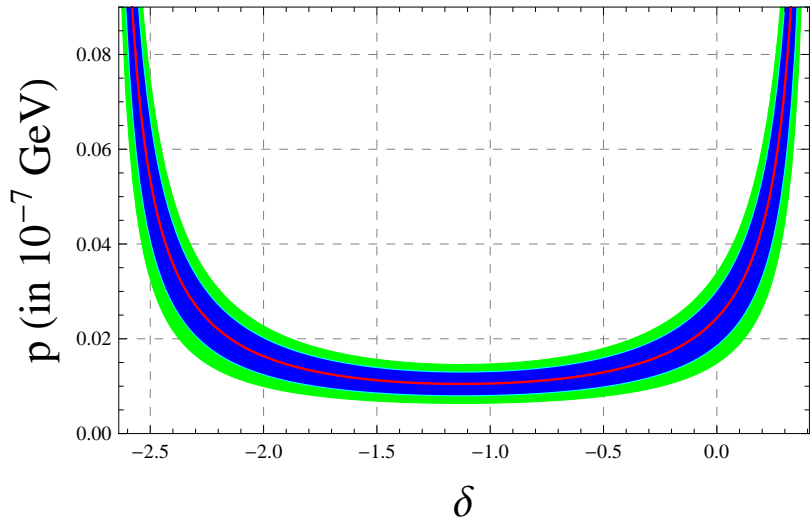}
\caption{$p\equiv |P_b|$ and $\delta$ allowed by the measured value of $\Delta A_{CP}$.
The (red) line represents central value, while inner (blue) and outer (green)
bands respectively represent 68\% confidence level (1$\sigma$) and 90\%
confidence level (1.64$\sigma$) regions based on error in $\Delta A_{CP}$.}
\label{fig:p-delta}
\bigskip
\end{center}
\end{figure}

Thus, given the values for
$|A_f|, \phi^f_T$ in Table \ref{tab:SCS} and $\gamma \simeq 70^\circ$~\cite{Charles:2011va}, 
all asymmetries depend on two unknown parameters, $|P_b|$ and $\delta$. Two of the
asymmetries, for  $D^+\to\pi^+\pi^0$ and $D^0\to K^0\bar K^0$, vanish because of the 
absence of a $P_b$ contribution in these processes. We will discuss these special cases below.
We note that the phases $\phi^f_T$ have a common sign ambiguity, $\phi^f_T \to - \phi^f_T$,
and the CP asymmetries (\ref{CPAsym}) are approximately invariant under a joint 
transformation $\phi^f_T \to -\phi^f_T\,,~\delta \to \pi - \delta$.

The averaged asymmetry difference $\Delta A_{CP}$ (\ref{AveAsym}) measured by 
LHCb and CDF 
may be used to constrain $|P_b|$ and $\delta$. The allowed values of $p\equiv |P_b|$ are 
plotted in Fig.~\ref{fig:p-delta} as a function of $\delta$ for a range of $\delta$,
--$2.64 \le \delta \le 0.41$, consistent with  (\ref{eqn:limits}). For a wide range of $\delta$
a CP-violating  penguin amplitude of magnitude $|P_b|= 0.01\times (10^{-7}~{\rm GeV})$ is
sufficient to account for the observed value of $\Delta A_{CP}$. This value corresponds to 
$|{\cal P + PA}|/|{\cal P}_b| \equiv (|P + PA|\lambda)/(|P_b|/|V^*_{cb}V_{ub}|)  \simeq 0.10$, 
measuring U-spin breaking in $P+PA$ as already shown in (\ref{SU3P+PA}).

%This is Figure 2
\begin{figure}
\begin{center}
\includegraphics[width=0.48\textwidth]{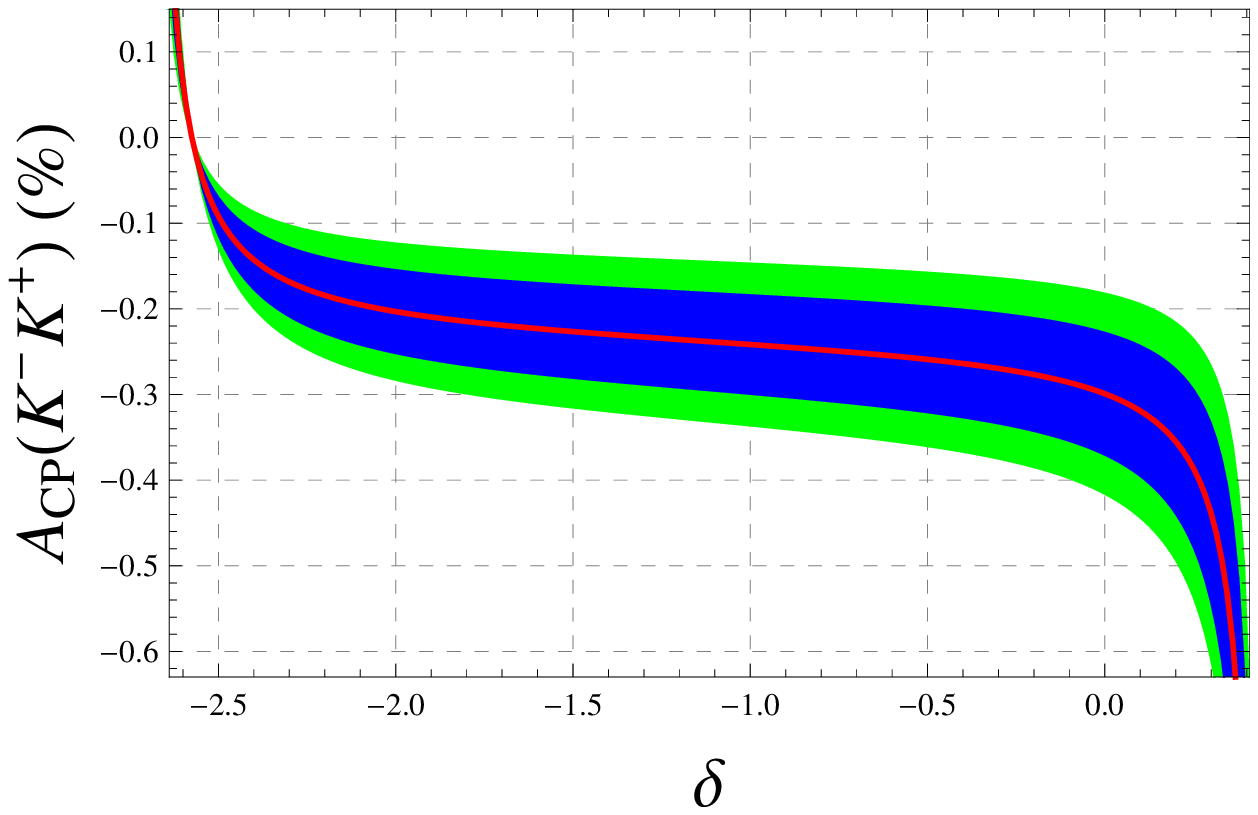}
\hskip 2mm
\includegraphics[width=0.48\textwidth]{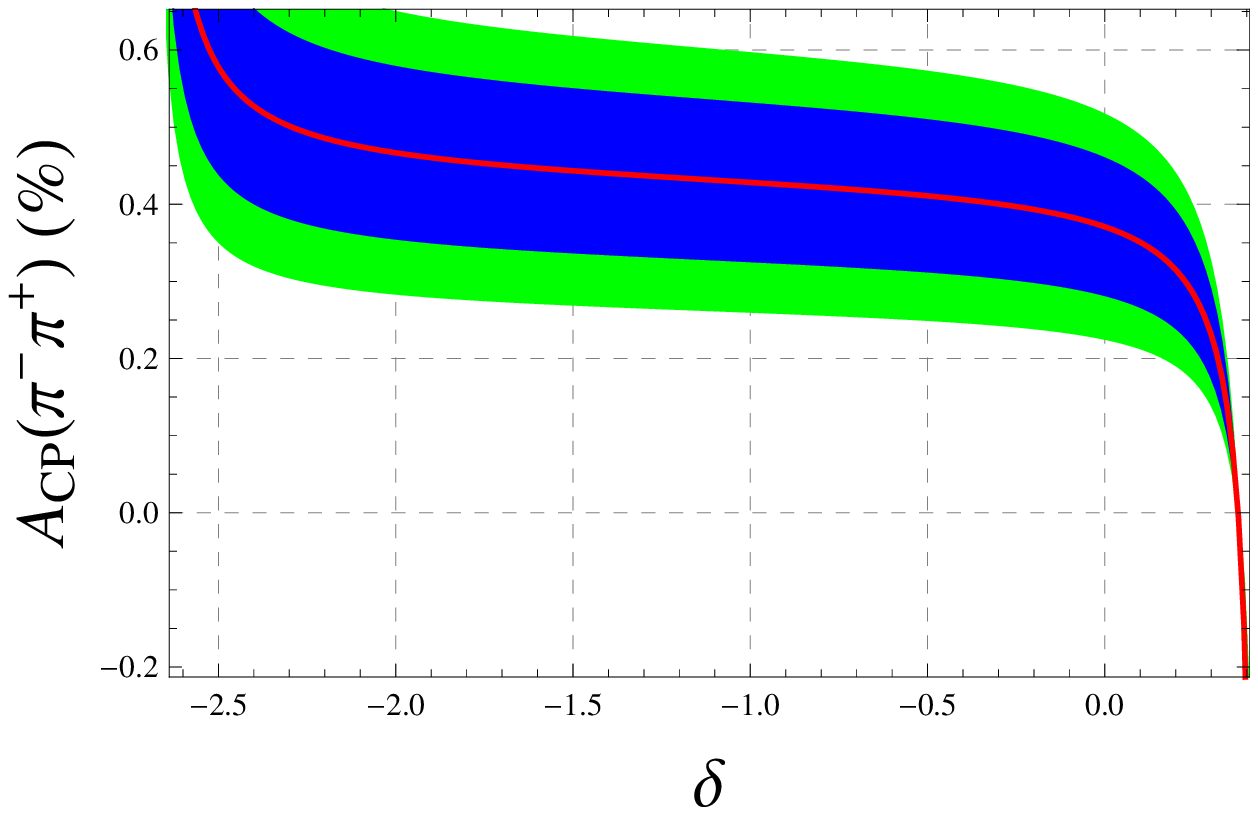}
\caption{$A_{CP}(D^0\to K^+K^-)$ and $A_{CP}(D^0\to\pi^+\pi^-)$ as a functions 
of the allowed values of $\delta$. The
(red) lines represent central values, while inner (blue) and outer (green)
bands respectively represent 68\% confidence level (1$\sigma$) and 90\%
confidence level (1.64$\sigma$) regions.}
\label{fig:KpmPpm}
\end{center}
\end{figure}
% This is Figure 3
\begin{figure}
\begin{center}
\includegraphics[width=0.48\textwidth]{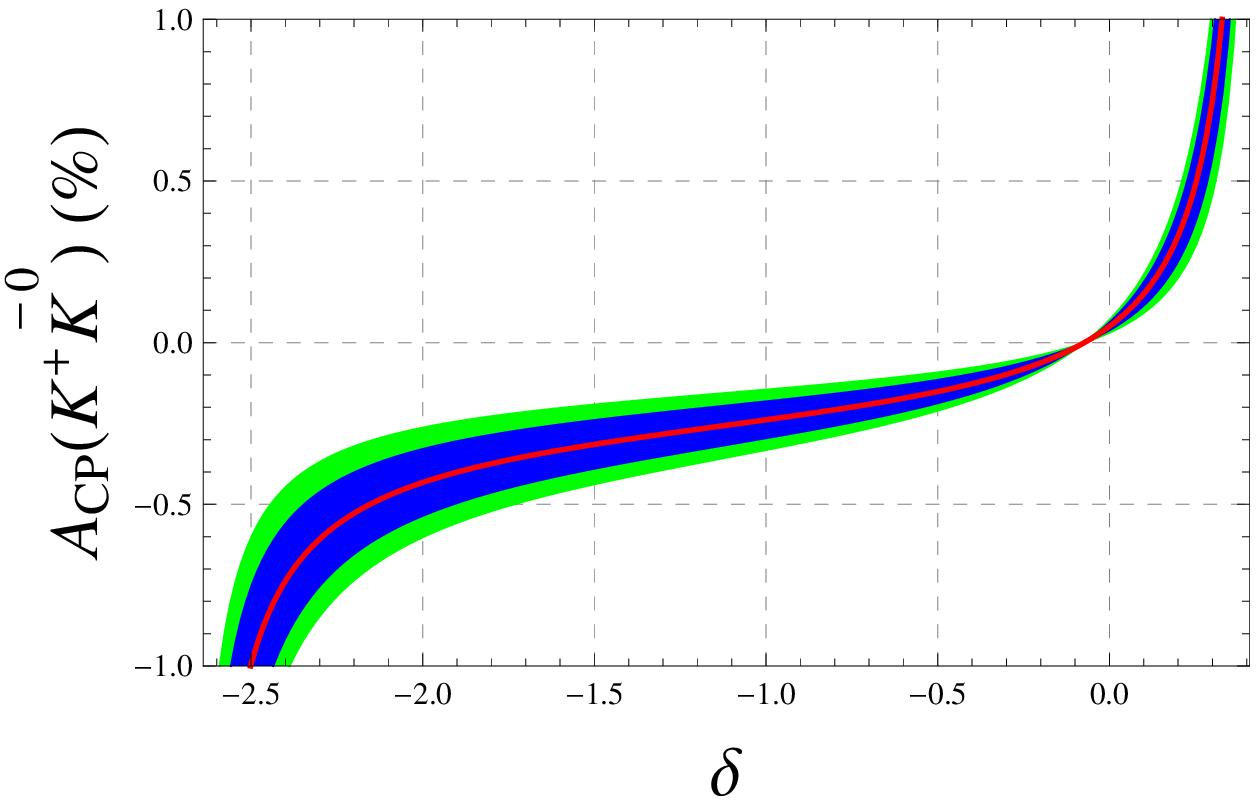}
\hskip 2mm
\includegraphics[width=0.48\textwidth]{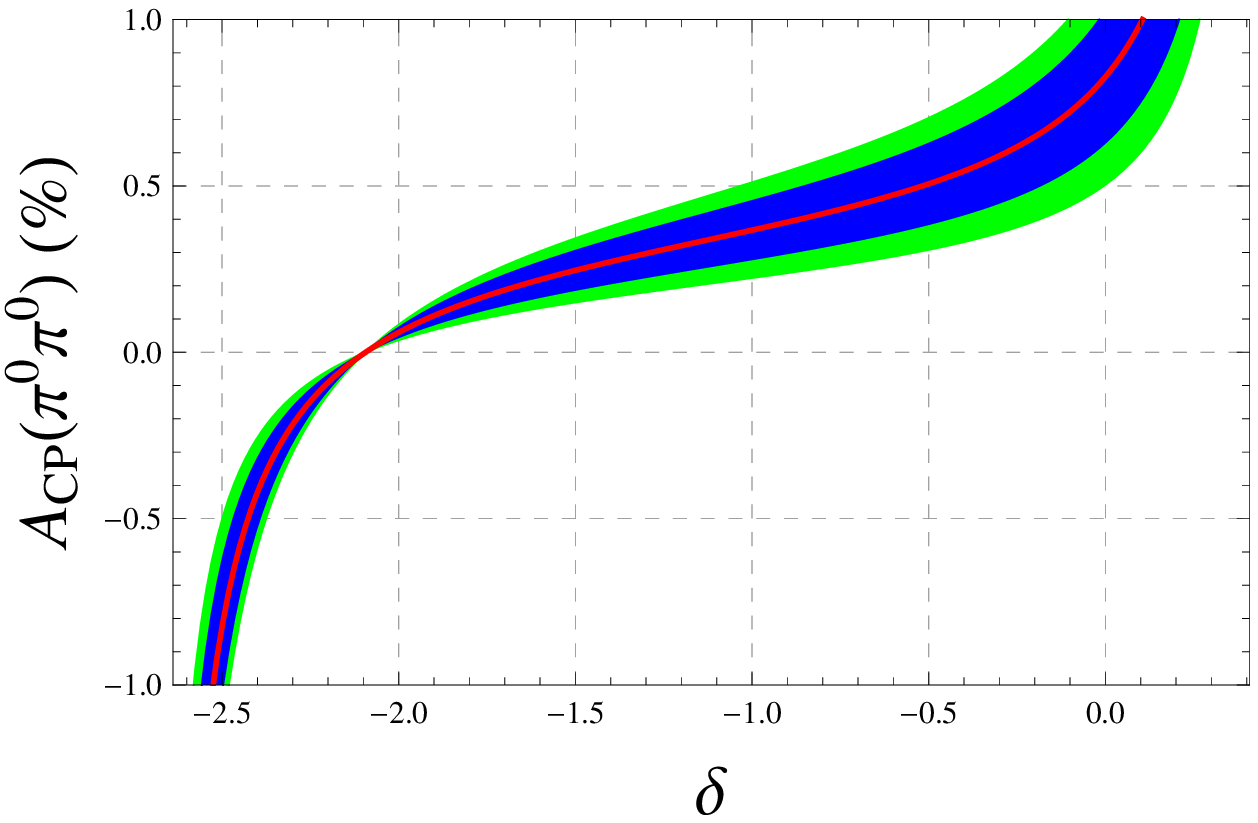}
\caption{$A_{CP}(D^+\to K^+\bar K^0)$ and $A_{CP}(D^0\to\pi^0\pi^0)$ as a functions 
of the allowed values of $\delta$. Color notations are as in Fig. \ref{fig:KpmPpm}.}
\label{fig:KpKbP0P0}
\end{center}
\end{figure}

In Fig\ \ref{fig:KpmPpm} we plot the asymmetries for $K^+K^-$ and $\pi^+\pi^-$ final states.
We note that for most values of $\delta$ one expects
\beq
A_{CP}(D^0\to K^+K^-) < 0\,,~~
A_{CP}(D^0\to\pi^+\pi^-) > 0\,,~~
\frac{|A_{CP}(D^0\to K^+K^-)|}{|A_{CP}(D^0\to\pi^+\pi^-)|} < 1~.
\eeqn
The central values (\ref{BoundsAsym}) measured by CDF support the signs 
of these two asymmetries.
More precise measurements of the two individual asymmetries  
are required  for testing their signs and can help pin down the unknown strong 
phase $\delta$, thereby predicting also the asymmetries in $D^0\to\pi^0\pi^0$ and
$D^+\to K^+\bar K^0$.

In Fig. \ref{fig:KpKbP0P0} we plot the predicted asymmetries for $D^+\to K^+\bar K^0$
and $D^0\to\pi^0\pi^0$. We note the correlation between these two asymmetries as
functions of $\delta$. The first asymmetry is negative for most values 
of $\delta$, while the second asymmetry is positive except for large negative values of 
$\delta$. 
Indeed, the central value of the current measurement~\cite{Ko:2010ng},
$A_{CP}(D^+\to K^+ \bar K^0)= (-0.16\pm 0.58\pm 0.25)\%$, is negative. The 
experimental error must be reduced in order to test the sign and to provide 
a useful constraint on $\delta$.
The property $A_{CP}(D^+\to K^+\bar K^0) < 0$ is based on assuming $PA_b=0$.
This assumption can be tested through the vanishing of $A_{CP}(D^0\to K^0\bar K^0)$,
as the absence of a $u$ quark in the final state forbids a $P_b$ contribution in this process.
The latter asymmetry may obtain a small contribution from an interference of $PA_b$ with an
SU(3)-breaking term in $E$~\cite{Cheng:2012xb}.

A quite general test of the Standard Model is $A_{CP}(D^+\to\pi^+\pi^0)=0$. The 
$\Delta I=1/2$ amplitude $P_b$ is absent in this process where the final
state has $I = 2$. Therefore one expects $A_{CP}(D^+\to \pi^+\pi^0)$ to vanish in the Standard 
Model at a very high precision. While higher order electroweak penguin amplitudes  involving
$\Delta I=3/2$ are enhanced in $B$ decays by the heavy $t$ quark mass~\cite{Gronau:1995hn}, such contributions are tiny in SCS charmed meson decays.

\section{Conclusion}

We conclude this talk by summarizing its main points:
\begin{itemize}
\item The  asymmetry difference, $\Delta A_{CP}\equiv A_{CP}(D^0\to K^+K^-)-
A_{CP}(D^0\to\pi^+\pi^-) \sim -0.7\%$, measured by the LHCb and CDF collaborations may 
originate in an enhancement from $1/m_c$ corrections of a CP-violating penguin amplitude 
$P_b$. These  corrections which are nonperturbative  are difficult to calculate reliably, 
reminiscent of the situation in $D^0$-$\bar D^0$ mixing. 
\item Experimental evidence supporting the above hypothesis exists in terms  of  large 
$1/m_c$ corrections causing an enhancement  of an exchange amplitude $E$ in CF decays.
\item A consistent broken flavor SU(3) framework is presented for CF and SCS decays of 
charmed mesons into two pseudoscalars, explaining in particular the large ratio
$|A(D^0\to K^+K^-)|/|A(D^0\to\pi^+\pi^-)| \approx 1.8$ and accounting quite precisely
for all four $D^0$ SCS decay rates. This scheme involves U-spin breaking 
in $\Delta U=1$ tree amplitudes given by ratios of meson decay constants and $D$ meson
form factors, and a purely U-spin breaking $\Delta U=0$ CP-conserving penguin amplitude
$P+PA$. 
\item Normalizing the value of $|P+PA|$ fitted by decay rates by the magnitude $|P_b|$ 
extracted from the measured $\Delta A_{CP}$, we find U-spin breaking in $P+PA$ to be  
about $10\%$ which seems reasonable.
\item This framework, which involves an unknown strong phase $\delta$, predicts CP asymmetries of order $({\rm several})\,\times 10^{-3}$ for $D^0\to \pi^+\pi^-, K^+K^-,  \pi^0\pi^0$ and $D^+\to K^+\bar K^0$, with specific signs for most value of $\delta$. 
Correlations as functions of $\delta$ are predicted between
$A_{CP}(D^0 \to \pi^+\pi^-)$ and $A_{CP}(D^0 \to K^+K^-)$ and between 
$A_{CP}(D^0 \to \pi^0\pi^0)$ and $A_{CP}(D^+ \to K^+\bar K^0)$.
\item Improving experimental limits on the individual asymmetries $A_{CP}(D^0 \to \pi^+\pi^-)$ 
and $A_{CP}(D^0 \to K^+K^-)$ would provide useful constraints on $\delta$.
\item Measuring a value for $A_{CP}(D^+\to \pi^+\pi^0)$ with magnitude larger than $0.1\%$ 
would provide unambiguous evidence for new physics due to $\Delta I=3/2$ operators.
\end{itemize}

In this talk we discussed charmed meson decays into two pseudoscalar mesons. 
One expects asymmetries of similar magnitudes in multi-body charm decays. 
Several decays into a vector meson and a pseudoscalar meson, in particular with 
charged particles in the final state, are experimentally feasible. This includes 
$D^+\to \phi (\to K^+K^-) \pi^+$ , for which an asymmetry $(0.51\pm 0.28\pm 0.05)\%$ 
has been published very recently~\cite{Staric:2011en} and reported 
at this conference \cite{Hayashii}. Such an asymmetry is due to 
interference of a color-suppressed tree amplitude and a flavor SU(3) singlet 
penguin amplitude involving $V^*_{cb}V_{ub}$ analogous to an amplitude ($s_P$) 
contributing to $B^+\to \phi\pi^+$~\cite{Dighe:1997wj}. 
Other  processes which are useful and should be studied are 
$D^+\to \rho^0(\to\pi^+\pi^-)\pi^+$, similar to $D^+\to \pi^+\pi^0$ but in which 
CP violation is permitted by isospin, and $D^+\to\bar K^{*0}(\to K^-\pi^+) K^+$ which 
resembles $D^+\to \bar K^0 K^+$ studied in this talk.  

It would  also be quite interesting to study CP asymmetries in a pair of processes, 
$D^0\to K_S(\to \pi^+\pi^-) K^{\pm}\pi^{\mp}$. These two asymmetries are expected 
to include contributions from CP violation in $K^0$-$\bar K^0$ mixing, similar to 
the situation in $D^+\to K_S\pi^+$
\cite{Bigi:1994aw,Grossman:2011zk,Ko:2010ng,delAmoSanchez:2011zza,Ko:2012pe}.

\end{document}

%% file: econfmacros.tex
\def\beq{\begin{equation}}
\def\eeq#1{\label{#1}\end{equation}}
\def\eeqn{\end{equation}}

\def \b{{\cal B}}

\def\beqa{\begin{eqnarray}}
\def\eeqa#1{\label{#1}\end{eqnarray}}
\def\eeqan{\end{eqnarray}}

\let\bar=\overbar

\def\Dslash{\not{\hbox{\kern-4pt $D$}}}
\def\dslash{\not{\hbox{\kern-2pt $\del$}}}

\def\msb{{\bar{\ssstyle M \kern -1pt S}}}

%% file: revTalk.bbl
\begin{thebibliography}{99}

\bibitem{Trabelsi} K. Trabelsi, these proceedings.

\bibitem{Schiller} M. T. Schiller, these proceedings.

\bibitem{Gronau:2007xg} 
  M.~Gronau,
  %``CP violation in beauty decays,''
  Int.\ J.\ Mod.\ Phys.\ A {\bf 22}, 1953 (2007)
  [arXiv:0704.0076 [hep-ph]].
  
  \bibitem{Buccella:1994nf} 
  F.~Buccella, M.~Lusignoli, G.~Miele, A.~Pugliese, and P.~Santorelli,
  %``Nonleptonic weak decays of charmed mesons,''
  Phys.\ Rev.\ D {\bf 51}, 3478 (1995)
  [hep-ph/9411286].
  
  \bibitem{Bianco:2003vb} 
  S.~Bianco, F.~L.~Fabbri, D.~Benson and I.~Bigi,
  %``A Cicerone for the physics of charm,''
  Riv.\ Nuovo Cim.\  {\bf 26N7}, 1 (2003)
  [hep-ex/0309021].
  
\bibitem{Grossman:2006jg} 
  Y.~Grossman, A.~L.~Kagan and Y.~Nir,
  %``New physics and CP violation in singly Cabibbo suppressed D decays,''
  Phys.\ Rev.\ D {\bf 75}, 036008 (2007)
  [hep-ph/0609178].
  
\bibitem{Bigi:2011re} 
  I.~I.~Bigi, A.~Paul and S.~Recksiegel,
  %``Conclusions from CDF Results on CP Violation in D^0 \to \pi^+\pi^-, K^+K^- and Future Tasks,''
  JHEP {\bf 1106}, 089 (2011)
  [arXiv:1103.5785 [hep-ph]].
  
  \bibitem{Golden:1989qx}
  M.~Golden and B.~Grinstein,
  %``Enhanced CP Violations in Hadronic Charm Decays,''
  Phys.\ Lett.\ B {\bf 222}, 501 (1989).
  %%CITATION = PHLTA,B222,501;%%
  
  \bibitem{Aaij:2011in}
  R.~Aaij {\it et al.} (LHCb Collaboration),
  %``Evidence for CP violation in time-integrated D0 -> h-h+ decay rates,''
%JR                        |||||||||||||||||
Phys.\ Rev.\ Lett.\ {\bf 108}, 111602 (2012).
  [arXiv:1112.0938 [hep-ex]]. 
  
  \bibitem{CDF}  T. Aaltonen {\it et al.} (CDF Collaboration), arXiv:1207.2158
  [hep-ex], submitted to Phys.\ Rev.\ Letters.
  
\bibitem {Vagnoni} V.M. Vagnoni, these proceedings.

 \bibitem{Bigi:2011em}
  I.~I.~Bigi and A.~Paul,
  %``On CP Asymmetries in Two-, Three- and Four-Body D Decays,''
  JHEP {\bf 1203}, 021 (2012)
 [ arXiv:1110.2862 [hep-ph]].
  %%CITATION = ARXIV:1110.2862;%%
  
   \bibitem{Isidori:2011qw} 
  G.~Isidori, J.~F.~Kamenik, Z.~Ligeti and G.~Perez,
  %``Implications of the LHCb Evidence for Charm CP Violation,''
  Phys.\ Lett.\ B {\bf 711}, 46 (2012)
  [arXiv:1111.4987 [hep-ph]].
  
  \bibitem{Brod:2011re}
  J.~Brod, A.~L.~Kagan and J.~Zupan,
  %``On the size of direct CP violation in singly Cabibbo-suppressed D decays,''
  arXiv:1111.5000 [hep-ph].

\bibitem{Pirtskhalava:2011va} 
  D.~Pirtskhalava and P.~Uttayarat,
  %``CP Violation and Flavor SU(3) Breaking in D-meson Decays,''
  Phys.\ Lett.\ B {\bf 712}, 81 (2012)
  [arXiv:1112.5451 [hep-ph]].
  
   \bibitem{Cheng:2012wr}
  H.~Y.~Cheng and C.~W.~Chiang,
  %``Direct CP violation in two-body hadronic charmed meson decays,''
%JR|||||||||||||||||||||||||||||||||||||
  Phys.\ Rev.\ D {\bf 85}, 034036 (2012) 
[arXiv:1201.0785 [hep-ph]].

\bibitem{Bhattacharya:2012ah} 
  B.~Bhattacharya, M.~Gronau and J.~L.~Rosner,
  %``CP asymmetries in singly-Cabibbo-suppressed $D$ decays to two pseudoscalar mesons,''
  Phys.\ Rev.\ D {\bf 85}, 054014 (2012)
  [arXiv:1201.2351 [hep-ph]].
  
  \bibitem{Li:2012cfa} 
  H.~N.~Li, C.~D.~Lu and F.~S.~Yu,
  %``Branching ratios and direct CP asymmetries in $D\to PP$ decays,''
  arXiv:1203.3120 [hep-ph].
  
  \bibitem{Franco:2012ck} 
  E.~Franco, S.~Mishima and L.~Silvestrini,
  %``The Standard Model confronts CP violation in $D^0 \to \pi^+\pi^-$ and $D^0 \to K^+K^-$,''
  JHEP {\bf 1205}, 140 (2012)
  [arXiv:1203.3131 [hep-ph]].

\bibitem{Brod:2012ud} 
  J.~Brod, Y.~Grossman, A.~L.~Kagan and J.~Zupan,
  %``A consistent picture for large penguins in D -> pi+ pi-, K+ K-,''
  arXiv:1203.6659 [hep-ph].

\bibitem{Cheng:2012xb} 
  H.~Y.~Cheng and C.~W.~Chiang,
  %``SU(3) symmetry breaking and CP violation in D -> PP decays,''
  arXiv:1205.0580 [hep-ph].
  
  \bibitem{Wang:2011uu} 
  K.~Wang and G.~Zhu,
  %``Can Up FCNC solve the $\Delta A_{CP}$ puzzle?,''
  Phys.\ Lett.\ B {\bf 709}, 362 (2012)
  [arXiv:1111.5196 [hep-ph]].
  
  \bibitem{Rozanov:2011gj} 
  A.~N.~Rozanov and M.~I.~Vysotsky,
  %``(\Delta A_{CP})_{LHCb} and the fourth generation,''
  arXiv:1111.6949 [hep-ph].
  
  \bibitem{Hochberg:2011ru} 
  Y.~Hochberg and Y.~Nir,
  %``Relating direct CP violation in D decays and the forward-backward asymmetry in $t\bar t$ production,''
  arXiv:1112.5268 [hep-ph].
  
  \bibitem{Giudice:2012qq} 
  G.~F.~Giudice, G.~Isidori and P.~Paradisi,
  %``Direct CP violation in charm and flavor mixing beyond the SM,''
  JHEP {\bf 1204}, 060 (2012)
  [arXiv:1201.6204 [hep-ph]].
  
  \bibitem{Altmannshofer:2012ur} 
  W.~Altmannshofer, R.~Primulando, C.~-T.~Yu and F.~Yu,
  %``New Physics Models of Direct CP Violation in Charm Decays,''
  JHEP {\bf 1204}, 049 (2012)
  [arXiv:1202.2866 [hep-ph]].
  
  \bibitem{Chen:2012am} 
  C.~H.~Chen, C.~Q.~Geng and W.~Wang,
  %``CP violation in $D^0 \to (K^- K^+, \pi^- \pi^+)$ from diquarks,''
  Phys.\ Rev.\ D {\bf 85}, 077702 (2012)
  [arXiv:1202.3300 [hep-ph]].
  
  \bibitem{Feldmann:2012js} 
  T.~Feldmann, S.~Nandi and A.~Soni,
  %``Repercussions of Flavour Symmetry Breaking on CP Violation in D-Meson Decays,''
  JHEP {\bf 1206}, 007 (2012)
  [arXiv:1202.3795 [hep-ph]].
  
  \bibitem{Hiller:2012wf} 
  G.~Hiller, Y.~Hochberg and Y.~Nir,
  %``Supersymmetric \Delta A_{CP},''
  arXiv:1204.1046 [hep-ph].
  
  \bibitem{Grossman:2012eb} 
  Y.~Grossman, A.~L.~Kagan and J.~Zupan,
  %``Testing for new physics in singly Cabibbo suppressed D decays,''
  arXiv:1204.3557 [hep-ph].
  
  \bibitem{Mannel:2012hb} 
  T.~Mannel and N.~Uraltsev,
  %``Charm CP Violation and the Electric Dipole Moments from the Charm Scale,''
  arXiv:1205.0233 [hep-ph].
 
 \bibitem{Chen:2012} C. H. Chen, C. Q. Qiang and W. Wang, arXiv:1206.5158 [hep-ph]. 
 
 \bibitem{average} http://www.slac.stanford.edu/xorg/hfag/charm/March12/DCPV/direct-indirect-cpv.html.
  
  \bibitem{E687} P. L. Frabetti {\it at al.} [E687 Collaboration], Phys.\ Rev.\  D {\bf 50}, 
  2693 (1994).

\bibitem{E791} E. M. Aitala et al. [E791 Collaboration], Phys. Lett. B
{\bf 421}, 405 (1998) [arXiv:hep-ex/9711003].

\bibitem{FOCUS} J. M. Link et al. [FOCUS Collaboration], Phys. Lett.
B {\bf 491}, 232 (2000) [Erratum-ibid. B {\bf 495}, 443 (2000)]
[arXiv:hep-ex/0005037].

\bibitem{CLEO} J. E. Bartelt et al. [CLEO Collaboration], Phys. Rev. D
{\bf 52}, 4860 (1995); S. E. Csorna et al. [CLEO Collaboration], Phys. Rev. D
{\bf 65}, 092001 (2002) [arXiv:hep-ex/0111024].

\bibitem{BaBar} B. Aubert et al. [BaBar Collaboration], Phys. Rev. Lett.
{\bf 100}, 061803 (2008) [arXiv:0709.2715 [hep-ex]].

\bibitem{Belle} M. Staric et al. [Belle Collaboration], Phys. Lett. B {\bf 670},
190 (2008) [arXiv:0807.0148 [hep-ex]].

\bibitem{Aaltonen:2011se} T. Aaltonen {\it et al.} (CDF Collaboration),
%``Measurement of CP--violating asymmetries in $D^0\to\pi^+\pi^-$ and $D^0\to K^+K^-$ decays at CDF,''
%JR|||||||||||||||||||||||||||||||||||||
  Phys.\ Rev.\ D {\bf 85}, 012009 (2012) [arXiv:1111.5023 [hep-ex]].
  
  \bibitem{Charles:2011va} 
  J.~Charles, O.~Deschamps, S.~Descotes-Genon, R.~Itoh, H.~Lacker, A.~Menzel, S.~Monteil and V.~Niess {\it et al.},
  %``Predictions of selected flavour observables within the Standard Model,''
  Phys.\ Rev.\ D {\bf 84}, 033005 (2011)
  [arXiv:1106.4041 [hep-ph]]. Updates available at http://ckmfitter.in2p3.fr.
  
   \bibitem{PDG} J. Beringer {\it et al.} (Particle Data Group),
  Phys.\ Rev.\ D {\bf 86}, 010001 (2012).
  
  \bibitem{Vainshtein:1975sv} 
  A.~I.~Vainshtein, V.~I.~Zakharov and M.~A.~Shifman,
  %``A Possible mechanism for the Delta T = 1/2 rule in nonleptonic decays of strange particles,''
  JETP Lett.\  {\bf 22}, 55 (1975)
  [Pisma Zh.\ Eksp.\ Teor.\ Fiz.\  {\bf 22}, 123 (1975)].
  
  \bibitem{Gronau:1994rj} 
  M.~Gronau, O.~F.~Hernandez, D.~London, and J.~L.~Rosner,
  %``Decays of B mesons to two light pseudoscalars,''
  Phys.\ Rev.\ D {\bf 50}, 4529 (1994)
  [hep-ph/9404283].
  
  %\cite{Bhattacharya:2008ss}
\bibitem{Bhattacharya:2008ss} 
  B.~Bhattacharya and J.~L.~Rosner,
  %``Flavor symmetry and decays of charmed mesons to pairs of light pseudoscalars,''
  Phys.\ Rev.\ D {\bf 77}, 114020 (2008)
  [arXiv:0803.2385 [hep-ph]].
  
  %\cite{Bhattacharya:2009ps}
\bibitem{Bhattacharya:2009ps} 
  B.~Bhattacharya and J.~L.~Rosner,
  %``Charmed meson decays to two pseudoscalars,''
  Phys.\ Rev.\ D {\bf 81}, 014026 (2010)
  [arXiv:0911.2812 [hep-ph]].
  
  \bibitem{Cheng:2010ry} 
  H.~Y.~Cheng and C.~W.~Chiang,
  %``Two-body hadronic charmed meson decays,''
  Phys.\ Rev.\ D {\bf 81}, 074021 (2010)
  [arXiv:1001.0987 [hep-ph]].
  
   \bibitem{Gronau:1999zt} 
  M.~Gronau,
  %``Resonant two-body D decays,''
  Phys.\ Rev.\ Lett.\  {\bf 83}, 4005 (1999)
  [hep-ph/9908237].
  
  \bibitem{Gronau:2000ru} 
  M.~Gronau and J.~L.~Rosner,
  %``U spin symmetry in doubly Cabibbo suppressed charmed meson decays,''
  Phys.\ Lett.\ B {\bf 500}, 247 (2001)
  [hep-ph/0010237].
  
  \bibitem{Bauer:1986bm} 
  M.~Bauer, B.~Stech and M.~Wirbel,
  %``Exclusive Nonleptonic Decays of D, D(s), and B Mesons,''
  Z.\ Phys.\ C {\bf 34}, 103 (1987).
  
  \bibitem{Neubert:1997uc} 
  M.~Neubert and B.~Stech,
  %``Nonleptonic weak decays of B mesons,''
  Adv.\ Ser.\ Direct.\ High Energy Phys.\  {\bf 15}, 294 (1998)
  [hep-ph/9705292].
  
  \bibitem{decay-constants}
  J. L. Rosner and S. Stone, ``Decay Constants of Charged Pseudoscalar Mesons",  
  mini-review in Ref.~\cite{PDG}.
  
  \bibitem{Besson:2009uv} 
  D.~Besson {\it et al.}  [CLEO Collaboration],
  %``Improved measurements of D meson semileptonic decays to pi and K mesons,''
  Phys.\ Rev.\ D {\bf 80}, 032005 (2009)
  [arXiv:0906.2983 [hep-ex]].
  
  \bibitem{Ko:2010ng} 
  B.~R.~Ko {\it et al.}  [Belle Collaboration],
  %``Search for CP violation in the decays $D^+_{(s)} \to K_S^0\pi^+$ and $D^+_{(s)} \to K_S^0K^+$,''
  Phys.\ Rev.\ Lett.\  {\bf 104}, 181602 (2010)
  [arXiv:1001.3202 [hep-ex]].
  
  \bibitem{Gronau:1995hn} 
  M.~Gronau, O.~F.~Hernandez, D.~London and J.~L.~Rosner,
  %``Electroweak penguins and two-body B decays,''
  Phys.\ Rev.\ D {\bf 52}, 6374 (1995)
  [hep-ph/9504327].
  
  \bibitem{Staric:2011en} 
  M.~Staric {\it et al.}  [Belle Collaboration],
  %``Search for CP Violation in $D^\pm$ Meson Decays to $\phi \pi^\pm$,''
  Phys.\ Rev.\ Lett.\  {\bf 108}, 071801 (2012)
  [arXiv:1110.0694 [hep-ex]].
  
  \bibitem{Hayashii} H. Hayashii, these proceedings.
  
  \bibitem{Dighe:1997wj} 
  A.~S.~Dighe, M.~Gronau and J.~L.~Rosner,
  %``B decays to charmless V P final states,''
  Phys.\ Rev.\ D {\bf 57}, 1783 (1998)
  [hep-ph/9709223].
  
  \bibitem{Bigi:1994aw} 
  I.~I.~Y.~Bigi and H.~Yamamoto,
  %``Interference between Cabibbo allowed and doubly forbidden transitions in D ---> K(S), K(L) + pi's decays,''
  Phys.\ Lett.\ B {\bf 349}, 363 (1995)
  [hep-ph/9502238].
  
  \bibitem{Grossman:2011zk} 
  Y.~Grossman and Y.~Nir,
  %``CP Violation in \tau ->\nu\pi K_S and D->\pi K_S: The Importance of K_S-K_L Interference,''
  JHEP {\bf 1204}, 002 (2012)
  [arXiv:1110.3790 [hep-ph]].
  
  \bibitem{Ko:2010ng} 
  B.~R.~Ko {\it et al.}  [Belle Collaboration],
  %``Search for CP violation in the decays $D^+_{(s)} \to K_S^0\pi^+$ and $D^+_{(s)} \to K_S^0K^+$,''
  Phys.\ Rev.\ Lett.\  {\bf 104}, 181602 (2010)
  [arXiv:1001.3202 [hep-ex]].
  
  \bibitem{delAmoSanchez:2011zza} 
  P.~del Amo Sanchez {\it et al.}  [BABAR Collaboration],
  %``Search for CP violation in the decay $D^\pm \to K_S^0\pi^\pm$,''
  Phys.\ Rev.\ D {\bf 83}, 071103 (2011)
  [arXiv:1011.5477 [hep-ex]].
  
  \bibitem{Ko:2012pe} 
  B.~R.~Ko {\it et al.}  [Belle Collaboration],
  %``Evidence for CP Violation in the Decay $D^+\rightarrow K^0_S\pi^+$,''
  arXiv:1203.6409 [hep-ex].
  
\end{thebibliography}
